\documentclass[onecolumn,american]{IEEEtran}
\usepackage[T1]{fontenc}
\usepackage[utf8]{inputenc}
\usepackage{amsmath}
\usepackage{amsthm}
\usepackage{amssymb}

\makeatletter
\theoremstyle{plain}
\newtheorem{thm}{\protect\theoremname}
\theoremstyle{definition}
\newtheorem{defn}[thm]{\protect\definitionname}
\theoremstyle{plain}
\newtheorem{prop}[thm]{\protect\propositionname}
\theoremstyle{plain}
\newtheorem{cor}[thm]{\protect\corollaryname}
\theoremstyle{remark}
\newtheorem{rem}[thm]{\protect\remarkname}

\makeatother

\usepackage{babel}
\providecommand{\corollaryname}{Corollary}
\providecommand{\definitionname}{Definition}
\providecommand{\propositionname}{Proposition}
\providecommand{\remarkname}{Remark}
\providecommand{\theoremname}{Theorem}

\begin{document}
\title{The Optimal Constant for One-Shot Channel Simulation and Strong Functional Representation Lemma}
\author{Cheuk Ting Li\\
Department of Information Engineering, The Chinese University of Hong Kong, Hong Kong, China \\
Email: ctli@ie.cuhk.edu.hk}
\maketitle
\begin{abstract}
We prove a tight bound for one-shot channel simulation via a stronger version of the strong functional representation lemma: for any random variables $X,Y$, there exists a random variable $Z$ with $I(Z;X)=H(Y|X,Z)=0$ and $H(Y|Z)\le I+\log I+(1-\gamma)\log e-\log\log e+O(I^{-1}\log I)$, where $I=I(X;Y)$ and $\gamma$ is the Euler-Mascheroni constant. A more convenient bound is $H(Y|Z)<I+\log(I+2)+1$. The first bound is notable in the sense that the constant term $(1-\gamma)\log e-\log\log e$ is optimal, and cannot be improved for any bound stated only in terms of $I$. Hence, this is the tightest form of the strong functional representation lemma up to the third order. We also prove that the Poisson functional representation is almost optimal up to a constant gap, and provide a simple finite-blocklength achievability bound that implies the $(1/2)\log n$ redundancy of the simulation of memoryless non-singular channel. These are proved via a generalization of the functional information, in particular, the exponential functional information, which closely approximates the cost of one-shot channel simulation.
\end{abstract}


\section{Introduction}

Channel simulation studies the simulation of a noisy channel using noiseless communication from an encoder to a decoder \cite{bennett2002entanglement,winter2002compression,cuff2013distributed,bennet2014reverse,yu2019exact,li2024channel}. Consider the one-shot variable-length channel simulation problem with unlimited common randomness \cite{harsha2010communication} (also see \cite{braverman2014public}), where we simulate the channel $P_{Y|X}$ using the common randomness $Z$. The communication cost (expected length of the variable-length codeword) is approximately the minimal conditional entropy of $Y$ given $Z$, defined as \cite{sfrl_trans}
\begin{equation}
H^{*}=H^{*}(X\to Y):=\inf_{P_{Z|X,Y}:\,I(Z;X)=H(Y|X,Z)=0}H(Y|Z).\label{eq:Hstar}
\end{equation}
The \emph{strong functional representation lemma} \cite{sfrl_trans} shows that $H^{*}\le I+\log(I+1)+4$ where $I=I(X;Y)$. Various improvements on this bound have been derived \cite{li2021unified,flamich2023adaptive,li2024pointwise,li2024channel,hill2026rejection,li2026discrete}, with the tightest one (for large $I$) being \cite{li2026discrete}:
\begin{equation}
H^{*}\le I+\log(I+5.22)+0.76.\label{eq:discrete_layered}
\end{equation}
Another notable result in \cite{hill2026rejection,flamich2025redundancy} shows that 
\begin{equation}
I_{\mathrm{F}}\le H^{*}\le I_{\mathrm{F}}+\log e,\label{eq:Hstar_IF}
\end{equation}
where $I_{\mathrm{F}}$ is the \emph{functional information} (which generalizes the excess functional information lower bound \cite{sfrl_trans,zhao2026rejection}) defined as
\[
I_{\mathrm{F}}=I_{\mathrm{F}}(X\to Y):=h(W|Y),
\]
where $W\in[0,\infty)$ is a continuous random variable with a conditional probability density function 
\begin{equation}
f_{W|Y}(w|y):=\mathbb{P}_{X\sim P_{X}}\big(w\le2^{\iota_{X;Y}(X;y)}\big)\label{eq:IF}
\end{equation}
given $Y$, $\iota_{X;Y}(x;y):=\log\frac{\mathrm{d}P_{X|Y}(x|y)}{\mathrm{d}P_{X}(x)}$ is the information density, and $h(W|Y)=\mathbb{E}[-\log f_{W|Y}(W|Y)]$ is the conditional differential entropy. Equivalently, $I_{\mathrm{F}}=h(W|Y)$ where $(X,Y)\sim P_{X,Y}$ and $W|(X,Y)\sim\mathrm{Unif}(0,2^{\iota_{X;Y}(X;Y)})$. This implies a bound $H^{*}\le I+\log(I+1)+2.45$ \cite{hill2026rejection}, which is not as tight as \cite{li2026discrete} for large $I$.

In this paper, we derive a new bound using the following quantity, which we call the \emph{exponential functional information}
\[
I_{\mathrm{E}}=I_{\mathrm{E}}(X\to Y):=h(V|Y),
\]
where $V\in[0,\infty)$ has a conditional probability density function 
\[
f_{V|Y}(v|y):=\mathbb{E}_{X\sim P_{X}}\big[e^{1-ve2^{-\iota_{X;Y}(X;y)}}\big]
\]
given $Y$. Equivalently, $I_{\mathrm{E}}=h(V|Y)$ where $(X,Y)\sim P_{X,Y}$ and $V|(X,Y)\sim\mathrm{Exp}(e2^{-\iota_{X;Y}(X;Y)})$. This can be considered as a smoothed version of the functional information \cite{hill2026rejection,flamich2025redundancy} in (\ref{eq:IF}). We use the exponential functional information to establish the following results.

\smallskip{}

\subsection*{Stronger strong functional representation lemma}

We show that
\begin{equation}
I_{\mathrm{E}}\le I_{\mathrm{F}}\le H^{*}\le I_{\mathrm{E}}+\log e\le I_{\mathrm{F}}+\log e.\label{eq:Is_IF}
\end{equation}
Moreover, we show in Theorem \ref{thm:PFR} that
\begin{align}
H^{*} & \le I_{\mathrm{E}}+\log e\nonumber \\
 & \le I+\left(1+\frac{\log e}{I+(1-\gamma)\log e}\right)\log\left(I+(2-\gamma)\log e\right)\nonumber \\
 & \quad\quad+(1-\gamma)\log e-\log\log e,\label{eq:Is_bound}
\end{align}
where $\gamma:=-\int_{0}^{\infty}e^{-t}\ln t\,\mathrm{d}t\approx0.577$ is the Euler-Mascheroni constant.\footnote{If entropy is in nats instead of bits, we have a simpler bound $H^{*}\le I+(1+\frac{1}{I+1-\gamma})\log(I+2-\gamma)+1-\gamma$, with a simpler constant term $1-\gamma$.}  For some simpler bounds,\footnote{The constants here are specific to base-$2$ entropy and log.}
\begin{equation}
H^{*}<I+\log(I+2)+1,\label{eq:Hstar_I_1}
\end{equation}
\begin{equation}
H^{*}<I+\log(I+4)+0.24.\label{eq:Hstar_I_2}
\end{equation}
The bound (\ref{eq:Hstar_I_2}) is stronger than all known bounds in terms of $I$ such as (\ref{eq:discrete_layered}), and can be applied to provide sharper bounds for one-shot lossy compression \cite{sfrl_trans,li2024pointwise}, rate-distortion-perception trade-off \cite{theis2021coding}, privacy-utility trade-off \cite{zamani2023privacy,zamani2026multi}, and oblivious relaying \cite{liu2026nonasymptotic}.

\smallskip{}

\subsection*{Exact characterization of the constant term of one-shot channel simulation}

The upper bound (\ref{eq:Is_bound}) is $I+\log I+(1-\gamma)\log e-\log\log e+O(I^{-1}\log I)$ for large $I$. We show in Theorem \ref{thm:example} that the constant term $(1-\gamma)\log e-\log\log e\le0.081183$ is the best possible constant, and cannot be improved for any bound that can be stated in terms of $I$ only. Hence, the bound (\ref{eq:Is_bound}) is third-order optimal.  This answers the question raised in \cite{li2026discrete}, and disproves the conjecture that the constant term is $-\log\log e$.

\smallskip{}

\subsection*{Almost-optimality of Poisson functional representation}

The bound $H^{*}\le I_{\mathrm{E}}+\log e$ is proved using the Poisson functional representation (PFR) construction \cite{sfrl_trans,li2021unified}, and hence PFR achieves a tighter upper bound than the ring toss code \cite{hill2026rejection} which has an upper bound $I_{\mathrm{F}}+\log e$ (\ref{eq:Hstar_IF}). PFR is almost optimal, in the sense that its conditional entropy is within $\log e+I_{\mathrm{E}}-I_{\mathrm{F}}\le\log e$ bits (or $1$ nat) from the optimum (\ref{eq:Is_IF}).

\smallskip{}

\subsection*{Finite-blocklength redundancy of memoryless channel simulation}

For the simulation of the memoryless channel $X^{n}\to Y^{n}$ where $(X_{i},Y_{i})\stackrel{\mathrm{iid}}{\sim}P_{X,Y}$ with variable-length codes, the exponential functional information can be used to give a finite-blocklength achievability bound for non-singular channels

\[
nI+\frac{\log n}{2}+\log\hat{S}+\frac{1}{2}\log\frac{2\pi e^{1-2\gamma}}{(\log e)^{2}}+\frac{\pi^{2}(\log e)^{3}}{12\hat{S}^{2}n},
\]
where $\hat{S}^{2}:=\mathbb{E}[\mathrm{Var}[\iota_{X;Y}(X;Y)|Y]]$ (Corollary \ref{cor:memoryless}). This applies to finite $n$ (even $n=1$). This refines upon the second-order asymptotic result in \cite{sriramu2024optimal,flamich2025redundancy} that the optimal cost is $nI+(1/2+o(1))\log n$, by giving an explicit upper bound on the constant term.\footnote{We study channel simulation with unlimited common randomness using a variable-length code for communication. If a fixed-length code is used instead, the redundancy is usually $O(\sqrt{n})$ \cite{cao2024channel}, instead of $(1/2+o(1))\log n$.}

\smallskip{}

\subsection*{Notations}

Entropy is in bits, and $\log$ is to the base $2$, though most results  should work as long as the bases of information measures (e.g., entropy, mutual information) and $\log$ are the same (unless otherwise specified). We consider general random variables lying in a Polish space with its Borel sigma algebra. For a random variable $X$, write $P_{X}$ for its probability measure, and $f_{X}$ for its probability density function (if it exists). Write $[n]:=\{1,\ldots,n\}$, $X^{n}=(X_{1},\ldots,X_{n})$. The differential entropy is $h(X):=\mathbb{E}[-\log f_{X}(X)]$, the differential min-entropy is $h_{\infty}(X):=-\log\sup_{x}f_{X}(x)$, and the information density is $\iota_{X;Y}(x;y):=\log\frac{\mathrm{d}P_{X|Y}(x|y)}{\mathrm{d}P_{X}(x)}$. The Euler-Mascheroni constant is $\gamma:=-\int_{0}^{\infty}e^{-t}\ln t\,\mathrm{d}t$.

\smallskip{}

\section{Generalized Functional Information }

We first introduce a class of information measures which generalizes the functional information in \cite{flamich2025redundancy,goc2024causal,hill2026rejection}. We also introduce an undirected variant.\footnote{Loosely speaking, these two quantities can be regarded as ``horizontal quantities'' in \cite{li2026discrete,hegazy2022randomized}.}

\smallskip{}

\begin{defn}
\label{def:gfi}Given a probability density function $f:[0,\infty)\to\mathbb{R}$ with finite differential entropy $h(f)$, the \emph{generalized functional information} from $X$ to $Y$ is\footnote{See the proof of Proposition \ref{prop:properties} for the proof that the two characterizations are equivalent.}
\begin{align*}
I_{f}(X\to Y) & :=h(V|Y)-h(f)\\
 & =I(X;\,Y,\,\iota(X;Y)+\log U),
\end{align*}
and the \emph{generalized undirected information} between $X,Y$ is
\begin{align*}
J_{f}(X;Y) & :=h(V)-h(f)\\
 & =I(X;Y)+I(X,Y;\,\iota(X;Y)+\log U),
\end{align*}
where $U\sim f$ is independent of $(X,Y)$, $V:=2^{\iota(X;Y)}U$, and we write $\iota(X;Y)=\iota_{X;Y}(X;Y)$.\footnote{If entropy and log are to the base $e$, we would have $V:=e^{\iota(X;Y)}U$ instead.} 
\end{defn}
\smallskip{}

We often write $I=I(X;Y)$, $I_{f}=I_{f}(X\to Y)$, $J_{f}=J_{f}(X;Y)$ if $X,Y$ are clear from the context. We are interested in these special cases:
\begin{itemize}
\item Mutual information can be recovered by taking $f$ to be $\mathrm{Lognormal}(0,\sigma^{2})$ with $\sigma^{2}\to\infty$, so $\log U\sim\mathrm{N}(0,(\sigma\log e)^{2})$, $I(X,Y;\,\iota(X;Y)+\log U)\to0$, and $I_{f},J_{f}\to I$.
\item $I_{\mathrm{F}}(X\to Y)=I_{f}(X\to Y)$ when $f$ is $\mathrm{Unif}(0,1)$, recovering the functional information \cite{flamich2025redundancy,goc2024causal,hill2026rejection}.
\item $I_{\mathrm{E}}(X\to Y):=I_{f}(X\to Y)$ when $f$ is $\mathrm{Exp}(e)$. We call this quantity the \emph{exponential functional information}.
\end{itemize}
We prove some properties of $I_{f}$ and $J_{f}$.

\smallskip{}

\begin{prop}
\label{prop:properties}The generalized functional/undirected information satisfies:
\begin{enumerate}
\item $I(X;Y)\le I_{f}(X\to Y)\le J_{f}(X;Y).$ 
\item $I(X;Y)=I_{f}(X\to Y)$ if and only if the channel $X\to Y$ is singular \cite{sriramu2024optimal} (i.e., $\iota_{X;Y}(X;Y)=\mathbb{E}[\iota_{X;Y}(X;Y)|Y]$ almost surely).\footnote{This generalizes \cite[Lemma 1]{hill2026rejection}.}
\item $I(X;Y)=J_{f}(X;Y)$ if and only if $\iota_{X;Y}(X;Y)$ is almost surely a constant.
\item $I_{f}(X\to Y)=0$ if and only if $J_{f}(X;Y)=0$, if and only if $I(X;Y)=0$.
\item \label{enu:properties_bd}Consider $f,g:[0,\infty)\to\mathbb{R}$ with finite differential entropies. Assume there exists a random variable $\Theta\in[0,\infty)$ such that if $U\sim f$ is independent of $\Theta$, then $\Theta U\sim g$.  We have
\[
I_{f}-I(\Theta;\Theta U)\le I_{g}\le I_{f},
\]
\[
J_{f}-I(\Theta;\Theta U)\le J_{g}\le J_{f}.
\]
In particular,
\[
I_{\mathrm{F}}-\gamma\log e\le I_{\mathrm{E}}\le I_{\mathrm{F}}.
\]
 
\end{enumerate}
\end{prop}
\begin{IEEEproof}
We first prove the equivalence of the different characterizations in Definition \ref{def:gfi}. By the change of variable formula $h(U)=h(\log U)+\mathbb{E}[\log U]-\log\log e$, we have
\begin{align*}
h(V)-h(U) & =h(\log V)+\mathbb{E}[\log V]-h(\log U)-\mathbb{E}[\log U]\\
 & =h(\iota(X;Y)+\log U)-h(\log U)\\
 & \quad\quad+\mathbb{E}[\iota(X;Y)+\log U]-\mathbb{E}[\log U]\\
 & =I(X,Y;\,\iota(X;Y)+\log U)+I.
\end{align*}
Also,
\begin{align*}
h(V|Y)-h(U) & =h(\log V|Y)+\mathbb{E}[\log V]-h(\log U)-\mathbb{E}[\log U]\\
 & =h(\iota(X;Y)+\log U|Y)-h(\log U)+I\\
 & =I(X;\,\iota(X;Y)+\log U\,|\,Y)+I\\
 & =I(X;\,Y,\,\iota(X;Y)+\log U).
\end{align*}
For the first property, we have $h(V)\ge h(V|Y)\ge h(V|X,Y)=\mathbb{E}[\iota(X;Y)]+h(f)=I(X;Y)+h(f)$. We have $h(V|Y)=h(V|X,Y)$ iff $I(X;V|Y)=0$, iff $I(X;\iota(X;Y)|Y)=0$, iff $\iota(X;Y)$ is almost surely a function of $Y$. We have $h(V)=h(V|X,Y)$ iff $I(V;X,Y)=0$, iff $I(\iota(X;Y);X,Y)=0$, iff $\iota(X;Y)$ is almost surely a constant.  For the fourth property, if $I_{f}(X\to Y)=0$, then $I(X;Y)=0$. If $I(X;Y)=0$, then $\iota(X;Y)=0$, so $h(V)=h(V|Y)=h(U|Y)=h(U)=h(f)$.  

For the last property, assuming $(X,Y)\sim P_{X,Y}$, $U\sim f$, $\Theta\sim P_{\Theta}$ are independent, since $X\to(Y,\iota(X;Y)+\log U)\to(Y,\iota(X;Y)+\log(\Theta U))$ forms a Markov chain, we have 
\begin{align*}
I_{g} & =I(X;\,Y,\,\iota(X;Y)+\log(\Theta U))\\
 & \le I(X;\,Y,\,\iota(X;Y)+\log U)\,=\,I_{f}.
\end{align*}
Similarly, $I(X,Y;\,\iota(X;Y)+\log(\Theta U))\le I(X,Y;\,\iota(X;Y)+\log U)$, so $J_{g}\le J_{f}$. Also, we have
\begin{align*}
 & I(\Theta;\Theta U)\\
 & =I(\Theta;\,\log(\Theta U))\\
 & =h(\log(\Theta U))-h(\log U)\\
 & =h(\log(\Theta U))+\mathbb{E}[\log(\Theta U)]-h(\log U)-\mathbb{E}[\log U]-\mathbb{E}[\log\Theta]\\
 & =h(\Theta U)-h(U)-\mathbb{E}[\log\Theta],
\end{align*}
where the last equality is by the change of variable formula. Hence,
\begin{align*}
I_{g} & =h(2^{\iota(X;Y)}\Theta U|Y)-h(\Theta U)\\
 & \ge h(2^{\iota(X;Y)}\Theta U|Y,\Theta)-h(\Theta U)\\
 & =h(2^{\iota(X;Y)}U|Y)+\mathbb{E}[\log\Theta]-h(\Theta U)\\
 & =I_{f}+\mathbb{E}[\log\Theta]+h(U)-h(\Theta U)\\
 & =I_{f}-I(\Theta;\Theta U).
\end{align*}
Similarly, $J_{g}\ge J_{f}-I(\Theta;\Theta U)$. 

Let $f$ be $\mathrm{Unif}(0,1)$, $g$ be $\mathrm{Exp}(e)$, and $\Theta\sim\mathrm{Gamma}(2,e)$, with $f_{\Theta}(\theta)=\theta e^{2-e\theta}$, $h(f)=h(g)=0$, and $\mathbb{E}[\log\Theta]=-\gamma\log e$. For $U\sim f$, we have $\mathbb{P}(\Theta U\ge t)=\int_{t}^{\infty}\theta e^{2-e\theta}(1-t/\theta)\mathrm{d}\theta=e^{-et}$, and hence $\Theta U\sim g$. We have $I_{\mathrm{F}}-\gamma\log e\le I_{\mathrm{E}}\le I_{\mathrm{F}}$.
\end{IEEEproof}
\smallskip{}

Proposition \ref{prop:properties} shows $I\le I_{f}\le J_{f}$. We then show that these three quantities are close within a logarithmic gap.

\smallskip{}

\begin{thm}
\label{thm:I_Jf}Consider $f:[0,\infty)\to\mathbb{R}$ with finite differential entropy. Writing $I=I(X;Y)$, $J_{f}=J_{f}(X;Y)$, $h=h(f)$, $h_{\infty}=h_{\infty}(f)=-\log\sup_{u}f(u)\le h(f)=0$, and $\eta:=\mathbb{E}_{U\sim f}[\log U]$, we have
\begin{equation}
I\ge J_{f}-\log\Big(\frac{J_{f}+h-h_{\infty}}{\log e}+1\Big)+h-\eta-\log e.\label{eq:I_Jf}
\end{equation}
Hence, the same bound applies to $I_{f}(X\to Y)$ in place of $J_{f}$ as well. In particular,
\begin{equation}
I\ge I_{\mathrm{F}}-\log\Big(\frac{I_{\mathrm{F}}}{\log e}+1\Big),\label{eq:I_IF}
\end{equation}
\begin{equation}
I\ge I_{\mathrm{E}}-\log\Big(\frac{I_{\mathrm{E}}}{\log e}+2\Big)+\gamma\log e.\label{eq:I_IE}
\end{equation}
\end{thm}
\smallskip{}

If an upper bound on $J_{f}$ in terms of $I(X;Y)$ is desired, we have the following corollary. Observe that $I_{\mathrm{E}}$ gives a smaller constant term than $I_{\mathrm{F}}$, which is a reason for $I_{\mathrm{E}}$ to give tighter bounds for one-shot results than $I_{\mathrm{F}}$.

\smallskip{}

\begin{cor}
\label{cor:Jf_I}Under the same assumptions as Theorem \ref{thm:I_Jf}, we have
\[
J_{f}\le I+\left(1+\frac{\log e}{I-h_{\infty}+\eta+\log e}\right)\log\left(\frac{I-h_{\infty}+\eta}{\log e}+2\right)-h+\eta+\log e.
\]
In particular,
\[
I_{\mathrm{F}}\le I+\left(1+\frac{\log e}{I}\right)\log\left(\frac{I}{\log e}+1\right),
\]
\[
I_{\mathrm{E}}\le I+\left(1+\frac{\log e}{I+(1-\gamma)\log e}\right)\log\left(\frac{I}{\log e}+2-\gamma\right)-\gamma\log e.
\]
\end{cor}
\smallskip{}

\begin{IEEEproof}
[Proof of Theorem \ref{thm:I_Jf} and Corollary \ref{cor:Jf_I}]We first prove the claim that if a random variable $W\in[0,\infty)$ has a probability density function satisfying $f_{W}(w)\le1$ for all $w$, and $\mathbb{E}[\ln W]\le b$, then $h(W)\le\psi^{-1}(b+1)\log e$, where $\psi^{-1}$ is the inverse function of $\psi(t):=t-\ln(t+1)$. To prove this claim, assume $t-\ln(t+1)=b+1$. Considering the cross entropy between $f_{W}$ and the probability density function $w\mapsto\min\{((t+1)w)^{-1-1/t},1\}$, we have
\begin{align*}
\frac{h(W)}{\log e} & \le-\int_{0}^{\infty}f_{W}(w)\ln\min\{((t+1)w)^{-1-1/t},1\}\mathrm{d}w\\
 & =\Big(1+\frac{1}{t}\Big)\int_{0}^{\infty}f_{W}(w)\max\{\ln((t+1)w),0\}\mathrm{d}w\\
 & \le\Big(1+\frac{1}{t}\Big)\Big(b+\ln(t+1)-\int_{0}^{1/(t+1)}f_{W}(w)\ln((t+1)w)\mathrm{d}w\Big)\\
 & \le\Big(1+\frac{1}{t}\Big)\Big(b+\ln(t+1)-\int_{0}^{1/(t+1)}\ln((t+1)w)\mathrm{d}w\Big)\\
 & =\Big(1+\frac{1}{t}\Big)\Big(b+\ln(t+1)+\frac{1}{t+1}\Big)\,=\,t.
\end{align*}
Let $\bar{f}:=\sup_{u}f(u)=2^{-h_{\infty}(f)}$. To prove the theorem, note that
\begin{align*}
\mathrm{sup}_{v}f_{V}(v) & \le\mathbb{E}\big[\mathrm{sup}_{v}f_{V|X,Y}(v|X,Y)\big]\\
 & =\mathbb{E}[2^{-\iota_{X;Y}(X;Y)}\bar{f}]\,\le\,\bar{f}.
\end{align*}
Hence, we can apply the claim on $W=\bar{f}V$ with $\sup_{w}f_{W}(w)\le1$ and 
\begin{align*}
\mathbb{E}[\ln W] & =\mathbb{E}[\ln(2^{\iota_{X;Y}(X;Y)}U\mathrm{sup}_{u}f(u))]\\
 & =\frac{I(X;Y)+\mathbb{E}[\log U]+\log\bar{f}}{\log e}.
\end{align*}
We obtain
\begin{align*}
h(V) & =h(W)-\log\bar{f}\le t\log e-\log\bar{f}=:s,
\end{align*}
where 
\[
t-\ln(t+1)=\frac{I(X;Y)+\mathbb{E}[\log U]+\log\bar{f}}{\log e}+1.
\]
Since $t=(s+\log\bar{f})/\log e$, we have
\begin{align*}
 & s-\log((s+\log\bar{f})/\log e+1)\\
 & =I(X;Y)+\mathbb{E}[\log U]+\log e.
\end{align*}
The result follows from $J_{f}\le s-h(f)$.

To prove the corollary, we will show that
\begin{equation}
\psi^{-1}(\tau)\le\tau+(1+\tau^{-1})\ln(\tau+1).\label{eq:psi_inv_bound}
\end{equation}
To show this, by the convexity of $\ln$,
\[
\ln(t+1)\le\ln(\tau+1)+\frac{t-\tau}{\tau+1}.
\]
Taking $t=\psi^{-1}(\tau)$, $t-\ln(t+1)=\tau$, we have
\[
t-\tau\le\ln(\tau+1)+\frac{t-\tau}{\tau+1},
\]
\[
t\le\tau+(1+\tau^{-1})\ln(\tau+1),
\]
which proves (\ref{eq:psi_inv_bound}). The corollary follows.

\end{IEEEproof}
\smallskip{}

We then give some more convenient bounds on $I_{\mathrm{E}}$.

\smallskip{}

\begin{cor}
\label{cor:Jf_I2}We have\footnote{The constants here are specific to base-$2$ entropy and log.}
\[
I_{\mathrm{E}}<I+\log(I+2)-0.6,
\]
\[
I_{\mathrm{E}}<I+\log(I+4)-1.21.
\]
\end{cor}
\smallskip{}

\begin{IEEEproof}
To prove $I_{\mathrm{E}}\le I+\log(I+b)+a$, applying $\tau\mapsto\tau+\log(\tau+b)+a$ to both sides of (\ref{eq:I_IE}), it suffices to prove that
\begin{align*}
t & \le t-\log\Big(\frac{t}{\log e}+2\Big)+\gamma\log e\\
 & \quad+\log\left(t-\log\Big(\frac{t}{\log e}+2\Big)+\gamma\log e+b\right)+a,
\end{align*}
for $t>0$. For any fixed $a,b$, this can be verified via interval arithmetic in a straightforward manner.
\end{IEEEproof}
\smallskip{}

We may bound $I_{f}$ and $J_{f}$ in terms of $I(X;Y)$ and the variance of the information density as well. The following result will be useful for giving a simple finite-blocklength bound which recovers the $(1/2)\log n$ redundancy term of asymptotic channel simulation \cite{sriramu2024optimal,flamich2025redundancy} (see Section \ref{sec:asymp}).

\smallskip{}

\begin{thm}
\label{thm:I_Jf_var}Consider $f:[0,\infty)\to\mathbb{R}$ with finite differential entropy. Writing  $I=I(X;Y)$, $S^{2}:=\mathrm{Var}[\iota_{X;Y}(X;Y)]$, $\hat{S}^{2}:=\mathbb{E}[\mathrm{Var}[\iota_{X;Y}(X;Y)|Y]]$, $\eta:=\mathbb{E}[\log U]$ and $\sigma^{2}:=\mathrm{Var}[\log U]$ for $U\sim f$, we have
\[
I_{f}\le I+\frac{1}{2}\log\Big(2\pi e\frac{\hat{S}^{2}+\sigma^{2}}{(\log e)^{2}}\Big)-h(f)+\eta,
\]
\[
J_{f}\le I+\frac{1}{2}\log\Big(2\pi e\frac{S^{2}+\sigma^{2}}{(\log e)^{2}}\Big)-h(f)+\eta.
\]
As a result,
\[
I_{\mathrm{F}}\le I+\frac{1}{2}\log\left(2\pi e\Big(\frac{\hat{S}^{2}}{(\log e)^{2}}+1\Big)\right)-\log e,
\]
\[
I_{\mathrm{E}}\le I+\frac{1}{2}\log\left(2\pi e\Big(\frac{\hat{S}^{2}}{(\log e)^{2}}+\frac{\pi^{2}}{6}\Big)\right)-(1+\gamma)\log e.
\]
\end{thm}
\smallskip{}

\begin{IEEEproof}
Since $V=2^{\iota_{X;Y}(X;Y)}U$, we have $\mathbb{E}[\log V]=I+\eta$, and
\begin{align*}
\mathrm{Var}[\log V] & =\mathrm{Var}[\iota_{X;Y}(X;Y)+\log U]\\
 & =S^{2}+\sigma^{2}.
\end{align*}
Since the log-normal distribution is the maximum entropy distribution given $\mathbb{E}[\log V]$ and $\mathrm{Var}[\log V]$, we have
\[
h(V)\le I+\eta+\frac{1}{2}\log\Big(2\pi e\frac{S^{2}+\sigma^{2}}{(\log e)^{2}}\Big).
\]
Also, $\mathbb{E}[\log V|Y]=\mathbb{E}[\iota(X;Y)|Y]+\eta$, and $\mathrm{Var}[\log V|Y]=\mathrm{Var}[\iota(X;Y)|Y]+\sigma^{2}$. Hence,
\begin{align*}
h(V|Y) & \le\mathbb{E}\left[\mathbb{E}[\iota(X;Y)|Y]+\eta+\frac{1}{2}\log\Big(2\pi e\frac{\mathrm{Var}[\iota(X;Y)|Y]+\sigma^{2}}{(\log e)^{2}}\Big)\right]\\
 & \le I+\eta+\frac{1}{2}\log\Big(2\pi e\frac{\hat{S}^{2}+\sigma^{2}}{(\log e)^{2}}\Big)
\end{align*}
\end{IEEEproof}
\smallskip{}

\begin{rem}
To further generalize Definition \ref{def:gfi}, we can define $J_{f}(X;Y\backslash Z):=h(V|Z)-h(f)$ for jointly-distributed $X,Y,Z$ (assume $U$ is independent of $X,Y,Z$). We have $J_{f}(X;Y)=J_{f}(X;Y\backslash\emptyset)$, $I_{f}(X\to Y)=J_{f}(X;Y\backslash Y)$ and $I(X;Y)=J_{f}(X;Y\backslash X,Y)$.
\end{rem}
\smallskip{}

\section{Bounding the Conditional Entropy of PFR}

We review the Poisson functional representation construction for one-shot channel simulation with unlimited common randomness \cite{sfrl_trans,li2021unified}. Let $Y_{1},Y_{2},\ldots\stackrel{\mathrm{iid}}{\sim}P_{Y}$ be independent of a Poisson point process $0\le T_{1}\le T_{2}\le\cdots$ of rate $1$. Take the common randomness to be $Z=(Y_{i},T_{i})_{i\in\mathbb{N}}$. Given $X\sim P_{X}$, we take $K:=\mathrm{argmin}_{k}2^{-\iota_{X;Y}(X;Y_{k})}T_{k}$, and $Y=Y_{K}$. It was shown in \cite{sfrl_trans} that $(X,Y)\sim P_{X,Y}$ and $\mathbb{E}[\log K]\le I(X;Y)+e^{-1}\log e+1$. This was improved to $\mathbb{E}[\log K]\le I(X;Y)+1$ in \cite{li2024pointwise}, which implies $H(Y|Z)\le H(K)\le I(X;Y)+\log(I(X;Y)+2)+2$.

We now prove a tighter bound 
\[
H(Y|Z)\le I+\log I+(1-\gamma)\log e-\log\log e+O(I^{-1}\log I).
\]
The bound is proved in the following theorem. The tightness of the constant $(1-\gamma)\log e-\log\log e$ will be established in Theorem \ref{thm:example}.

\smallskip{}

\begin{thm}
\label{thm:PFR}The random variable $Z$ given by the Poisson functional representation satisfies $I(Z;X)=H(Y|X,Z)=0$ and
\[
H^{*}(X\to Y)\le H(Y|Z)\le I_{\mathrm{E}}(X\to Y)+\log e.
\]
As a result, writing $H^{*}=H^{*}(X\to Y)$, $I=I(X;Y)$, $c:=(1-\gamma)\log e$, 
\begin{align*}
H^{*} & \le I+\left(1+\frac{\log e}{I+c}\right)\log\left(I+c+\log e\right)+c-\log\log e.
\end{align*}
For more convenient bounds,\footnote{The constants here are specific to base-$2$ entropy and log.}
\[
H^{*}<I+\log(I+2)+1,
\]
\[
H^{*}<I+\log(I+4)+0.24.
\]
\end{thm}
\smallskip{}

\begin{IEEEproof}
Let $\ell(\tau):=-\tau\log\tau$. Write $\iota(x;y)=\iota_{X;Y}(x;y)$. We have
\begin{align*}
 & H(Y|Z)\\
 & \le H(K|Z)\\
 & =\mathbb{E}_{Z}\bigg[\sum_{k=1}^{\infty}\ell(P_{K|Z}(k|Z))\bigg]\\
 & =\mathbb{E}_{Z}\bigg[\sum_{k=1}^{\infty}\ell\Big(P_{X}\big(\big\{ x:2^{-\iota(x;Y_{k})}T_{k}\le\min_{i}2^{-\iota(x;Y_{i})}T_{i}\big\}\big)\Big)\bigg]\\
 & \stackrel{(a)}{=}\int\int_{0}^{\infty}\mathbb{E}_{Z}\bigg[\ell\Big(P_{X}\big(\big\{ x:2^{-\iota(x;y)}t\le\min_{i}2^{-\iota(x;Y_{i})}T_{i}\big\}\big)\Big)\bigg]\,\mathrm{d}t\,P_{Y}(\mathrm{d}y)\\
 & \stackrel{(b)}{\le}\int\int_{0}^{\infty}\ell\bigg(\mathbb{E}_{Z}\bigg[P_{X}\big(\big\{ x:2^{-\iota(x;y)}t\le\min_{i}2^{-\iota(x;Y_{i})}T_{i}\big\}\big)\bigg]\bigg)\,\mathrm{d}t\,P_{Y}(\mathrm{d}y)\\
 & =\int\int_{0}^{\infty}\ell\bigg(\mathbb{P}_{X,Z}\Big(2^{-\iota(X;y)}t\le\min_{i}2^{-\iota(X;Y_{i})}T_{i}\Big)\bigg)\,\mathrm{d}t\,P_{Y}(\mathrm{d}y)\\
 & \stackrel{(c)}{=}\int\int_{0}^{\infty}\ell\Big(\mathbb{E}_{X}\big[\exp(-2^{-\iota(X;y)}t)\big]\Big)\,\mathrm{d}t\,P_{Y}(\mathrm{d}y)\\
 & \stackrel{(d)}{=}\int\int_{0}^{\infty}\ell\Big(\mathbb{E}_{X}\big[\exp(1-e2^{-\iota(X;y)}t)\big]\Big)\,\mathrm{d}t\,P_{Y}(\mathrm{d}y)+\log e\\
 & =I_{\mathrm{E}}(X\to Y)+\log e,
\end{align*}
where (a) is by the Mecke equation \cite[Theorem 4.1]{last2017lectures} applied to the Poisson point process $Z=(Y_{i},T_{i})_{i\in\mathbb{N}}$ with intensity measure $P_{Y}\times\lambda_{[0,\infty)}$ (where $\lambda_{[0,\infty)}$ is the Lebesgue measure over $[0,\infty)$), (b) is by Jensen's inequality, (c) is because $\min_{i}2^{-\iota(x;Y_{i})}T_{i}\sim\mathrm{Exp}(\int2^{\iota(x;y)}P_{Y}(\mathrm{d}y))=\mathrm{Exp}(1)$, and (d) is by $h(eV)=h(V)+\log e$.  The other bounds follow from Corollary \ref{cor:Jf_I} and Corollary \ref{cor:Jf_I2}.
\end{IEEEproof}
\smallskip{}

\section{Memoryless Channel Simulation}\label{sec:asymp}

In this section, we consider the simulation of the memoryless channel $P_{Y|X}$ with i.i.d. input $X^{n}=(X_{1},\ldots,X_{n})$, i.e., we take $X=X^{n}$, $Y=Y^{n}$ where $(X_{i},Y_{i})\stackrel{\mathrm{iid}}{\sim}P_{X,Y}$. The second-order term of the communication cost as $n\to\infty$ has been characterized in \cite{sriramu2024optimal} (under some restrictions), and it has been established in \cite{flamich2025redundancy} that $I_{\mathrm{F}}(X^{n}\to Y^{n})=nI(X;Y)+(1/2+o(1))\log n$ for non-singular channels (i.e., $\mathrm{Var}[\iota_{X;Y}(X;Y)|Y]>0$). The case for singular channels have also been studied in \cite{flamich2026singular}.

We now prove a finite-blocklength achievability bound in terms of only the mutual information and the conditional variance of the information density, which is strong enough to achieve the second-order term $(1/2)\log n$ for non-singular channels. Unlike previous results \cite{sriramu2024optimal,flamich2025redundancy}, our bound applies to finite $n$. 

\smallskip{}

\begin{cor}
\label{cor:memoryless}Let $I=I(X;Y)$, $\hat{S}^{2}:=\mathbb{E}[\mathrm{Var}[\iota_{X;Y}(X;Y)|Y]]$. The communication cost $H^{*}=H^{*}(X^{n}\to Y^{n})$ of simulating the channel $X^{n}\to Y^{n}$ (\ref{eq:Hstar}) is upper-bounded by
\begin{align}
H^{*} & \le nI+\frac{1}{2}\log\left(2\pi e\Big(\frac{n\hat{S}^{2}}{(\log e)^{2}}+\frac{\pi^{2}}{6}\Big)\right)-\gamma\log e.\label{eq:memoryless}
\end{align}
Hence, for non-singular channels ($\hat{S}^{2}>0$),
\begin{align}
H^{*} & \le nI+\frac{\log n}{2}+\log\hat{S}+\frac{1}{2}\log\frac{2\pi e^{1-2\gamma}}{(\log e)^{2}}+\frac{\pi^{2}(\log e)^{3}}{12\hat{S}^{2}n}.\label{eq:memoryless_nonsing}
\end{align}
For singular channels ($\hat{S}^{2}=0$), 
\begin{equation}
H^{*}\le nI+\frac{1}{2}\log\frac{2\pi^{3}e^{1-2\gamma}}{6}.\label{eq:memoryless_sing}
\end{equation}
\end{cor}
\smallskip{}

\begin{IEEEproof}
(\ref{eq:memoryless}) is an immediate consequence of the bounds about $I_{\mathrm{E}}$ in Theorems \ref{thm:I_Jf_var} and \ref{thm:PFR}. (\ref{eq:memoryless_nonsing}) follows from (\ref{eq:memoryless}) and $\log(a+b)\le\log a+(b/a)\log e$.
\end{IEEEproof}
\smallskip{}

\section{Tightness of the Strong Functional Representation Lemma}

Recall that Theorem \ref{thm:PFR} shows that, for large $I$, 
\[
H^{*}\le I+\log I+(1-\gamma)\log e-\log\log e+O(I^{-1}\log I).
\]
We can raise the question whether this bound is the tightest possible bound in terms of $I$, or we can find another tighter bound $H^{*}\le g(I)$ for some function $g$. The following theorem shows that there exist discrete random variables $X,Y$ where 
\[
H^{*}\ge I+\log I+(1-\gamma)\log e-\log\log e-O(I^{-1}(\log I)^{2}),
\]
coinciding with the upper bound for the three most significant terms. The constant term $(1-\gamma)\log e-\log\log e$ is optimal, which answers the question raised in \cite{li2026discrete} on the optimal constant. As a result,
\begin{align*}
 & \sup_{P_{X,Y}:\,I(X;Y)=I}H^{*}(X\to Y)\\
 & =I+\log I+(1-\gamma)\log e-\log\log e+O(I^{-1}(\log I)^{2}).
\end{align*}

\smallskip{}

\begin{thm}
\label{thm:example}For $I>2\log e$,
\begin{align*}
 & \sup_{P_{X,Y}:\,I(X;Y)=I}H^{*}(X\to Y)\\
 & >I+\log I+(1-\gamma)\log e-\log\log e-12\frac{(\log e)^{2}}{I}\Big(\ln\frac{I}{\log e}\Big)^{2},
\end{align*}
where the supremum is over finite discrete joint distributions $P_{X,Y}$ with $I(X;Y)=I$.
\end{thm}
\smallskip{}

\begin{IEEEproof}
Let $0<\alpha\le1/2$, and $\Theta\in[0,1]$ be a beta random variable with $f_{\Theta}(\theta)=\alpha\theta^{\alpha-1}$. Conditional on $\Theta$, we generate $\bar{X}\subseteq\mathbb{N}$ to be a random subset such that $\mathbf{1}\{i\in\bar{X}\}\stackrel{\mathrm{iid}}{\sim}\mathrm{Bern}(\Theta)$ for $i\in\mathbb{N}$ conditional on $\Theta$, i.e., we include each number with probability $\Theta$. Fix $k\in\mathbb{N}$. Let $X:=\bar{X}\cap[k]$ if $\bar{X}\cap[k]\neq\emptyset$, or $X:=[k]$ if $\bar{X}\cap[k]=\emptyset$. Conditional on $X$, generate $Y\in X$ uniformly at random.  By Chebyshev's inequality, for $\delta>0$, $\mathbb{P}(|X|/k\le\Theta-\delta\,|\,\Theta)\le1/(k\delta^{2})$ almost surely. Taking $\delta=k^{-1/3}$, 
\begin{align*}
I(X;Y) & =H(Y)-H(Y|X)\\
 & =\mathbb{E}[-\log(|X|/k)]\\
 & \le(\log k)\mathbb{P}(|X|/k\le\Theta-k^{-1/3})+\mathbb{E}[\min\{\log k,-\log(\Theta-k^{-1/3})\}]\\
 & \le k^{-1/3}\log k+\mathbb{E}[\min\{\log k,-\log(\Theta-k^{-1/3})\}].
\end{align*}
Hence, 
\begin{equation}
\mathrm{limsup}_{k}I(X;Y)\le\mathbb{E}[-\log\Theta]=\alpha^{-1}\log e.\label{eq:example_limsup}
\end{equation}
 We then prove that
\begin{equation}
H^{*}(X\to Y)=H(\min X).\label{eq:example_Hstar}
\end{equation}
For the achievability, take $Z:[k]\to[k]$ to be a uniformly random permutation, and $Y=\mathrm{argmin}_{i\in X}Z(i)$. By symmetry, $H(Y|Z)=H(Y|Z=(i\mapsto i))=H(\min X)$. For the converse, consider any $Z$ with $I(X;Z)=0$. For $z$ where $P_{X|Z=z}=P_{X}$ (which holds for $P_{Z}$-almost every $z$), for every $A\in[k]$, by symmetry,
\begin{align*}
\mathbb{P}(Y\in A|Z=z) & \le\mathbb{P}(X\cap A\neq\emptyset|Z=z)\\
 & =\mathbb{P}(X\cap A\neq\emptyset)\\
 & =\mathbb{P}(X\cap[|A|]\neq\emptyset)\\
 & =\mathbb{P}(\min X\le|A|).
\end{align*}
Hence, $P_{Y|Z=z}$ is majorized by $P_{\min X}$. Since entropy is Schur concave, $H(Y|Z)\ge H(\min X)$. This completes the proof of (\ref{eq:example_Hstar}).

Let $M:=\min\bar{X}$. Since $P_{\min X}\to P_{M}$ pointwise as $k\to\infty$ (due to $\mathbb{P}(\bar{X}\cap[k]\neq\emptyset)\to1$), we have 
\begin{equation}
\mathrm{liminf}_{k}\,H(\min X)\ge H(M).\label{eq:example_HminX_HM}
\end{equation}
We now lower-bound $H(M)$. We have $M|\Theta\sim\mathrm{Geom}(\Theta)$, and hence $M$ follows a mixture of geometric distributions. Direct evaluation gives
\[
P_{M}(m)=\frac{\alpha}{m}\prod_{i=1}^{m}\Big(1+\frac{\alpha}{i}\Big)^{-1}.
\]
Hence,
\begin{align}
H(M) & =\mathbb{E}\left[\log M+\sum_{i=1}^{M}\log\Big(1+\frac{\alpha}{i}\Big)\right]-\log\alpha\nonumber \\
 & \stackrel{(a)}{\ge}\mathbb{E}\left[\log M+\sum_{i=1}^{M}\Big(\frac{\alpha}{i}-\frac{\alpha^{2}}{2i^{2}}\Big)\log e\right]-\log\alpha\nonumber \\
 & \stackrel{(b)}{\ge}\mathbb{E}[\log M]+\alpha\mathbb{E}\bigg[\sum_{i=1}^{M}\frac{1}{i}\bigg]\log e-\frac{\alpha^{2}\pi^{2}}{12}\log e-\log\alpha\nonumber \\
 & \stackrel{(c)}{\ge}\mathbb{E}[\log M]+\alpha\mathbb{E}[\ln M+\gamma]\log e-\frac{\alpha^{2}\pi^{2}}{12}\log e-\log\alpha\nonumber \\
 & =(1+\alpha)\mathbb{E}[\log M]-\log\alpha+\alpha\Big(\gamma-\frac{\alpha\pi^{2}}{12}\Big)\log e,\label{eq:example_HM}
\end{align}
where (a) is due to $\ln(1+t)\ge t-t^{2}/2$ for $t\ge0$, (b) is by $\sum_{i=1}^{\infty}i^{-2}=\pi^{2}/6$, and (c) is due to $\sum_{i=1}^{M}i^{-1}=\int_{1}^{M+1}\lfloor t\rfloor^{-1}\mathrm{d}t\ge\int_{1}^{\infty}(\lfloor t\rfloor^{-1}-t^{-1})\mathrm{d}t+\int_{1}^{M+1}t^{-1}\mathrm{d}t=\gamma+\ln(M+1)$. To bound $\mathbb{E}[\log M]$, consider $T\sim\mathrm{Exp}(\beta)$. We have $\lceil T\rceil\sim\mathrm{Geom}(1-e^{-\beta})$, and $\mathbb{E}[\log\lceil T\rceil]\ge\mathbb{E}[\log T]=-\log\beta-\gamma\log e$. Hence, 
\begin{align*}
 & \mathbb{E}[\log M|\Theta]\ge-\log\left(-\ln(1-\Theta)\right)-\gamma\log e.
\end{align*}
Therefore, for $\delta=\alpha/e<1/e$,
\begin{align*}
 & \mathbb{E}[\ln M]\\
 & \ge\mathbb{E}[\mathbb{E}[\ln M|\Theta]\cdot\mathbf{1}\{\Theta\le\delta\}]\\
 & \ge\mathbb{E}\left[-\ln\left(-\ln(1-\Theta)\right)\cdot\mathbf{1}\{\Theta\le\delta\}\right]-\gamma\\
 & \ge\mathbb{E}\left[-\ln\left(-\frac{\Theta}{\delta}\ln(1-\delta)\right)\cdot\mathbf{1}\{\Theta\le\delta\}\right]-\gamma\\
 & \ge\mathbb{E}[(-\ln\Theta)\cdot\mathbf{1}\{\Theta\le\delta\}]-\ln\Big(-\frac{\ln(1-\delta)}{\delta}\Big)-\gamma\\
 & =\frac{e^{\alpha\ln\delta}(1-\alpha\ln\delta)}{\alpha}-\ln\Big(-\frac{\ln(1-\delta)}{\delta}\Big)-\gamma\\
 & \ge\frac{(1+\alpha\ln\delta)(1-\alpha\ln\delta)}{\alpha}-\delta-\gamma\\
 & =\frac{1}{\alpha}-\alpha(\ln\delta)^{2}-\delta-\gamma\\
 & =\frac{1}{\alpha}-\alpha(\ln\alpha-1)^{2}-\frac{\alpha}{e}-\gamma\\
 & \ge\frac{1}{\alpha}-\frac{9}{4}.
\end{align*}
Substituting into (\ref{eq:example_HM}),
\begin{align*}
\frac{H(M)}{\log e} & \ge(1+\alpha)\left(\frac{1}{\alpha}-\alpha(\ln\alpha-1)^{2}-\frac{\alpha}{e}-\gamma\right)-\ln\alpha+\alpha\Big(\gamma-\frac{\alpha\pi^{2}}{12}\Big)\\
 & \ge\frac{1}{\alpha}-\alpha(\ln\alpha-1)^{2}-\frac{\alpha}{e}-\gamma+\alpha\left(\frac{1}{\alpha}-\frac{9}{4}\right)-\ln\alpha+\alpha\Big(\gamma-\frac{\alpha\pi^{2}}{12}\Big)\\
 & =\frac{1}{\alpha}-\ln\alpha+1-\gamma-\alpha\left((\ln\alpha-1)^{2}+\frac{1}{e}+\frac{9}{4}-\gamma+\frac{\alpha\pi^{2}}{12}\right)\\
 & >\frac{1}{\alpha}-\ln\alpha+1-\gamma-12\alpha(\ln\alpha)^{2},
\end{align*}
where the last inequality holds for $\alpha\le1/2$. Consider $I>2\log e$ and small $\epsilon>0$. From (\ref{eq:example_limsup}), taking $\alpha=(\log e)/(I-\epsilon)\le1/2$, we have $I(X;Y)\le I$ for large enough $k$. By (\ref{eq:example_Hstar}) and (\ref{eq:example_HminX_HM}), $H^{*}(X\to Y)=H(\min X)\ge H(M)-\epsilon$ for large enough $k$. Letting $W$ be a discrete random variable with $I(W;X,Y)=0$ and $H(W)=I-I(X;Y)$, we have $I(X,W;Y,W)=I$ and $H^{*}((X,W)\to(Y,W))\ge H^{*}(X\to Y)$. Hence, taking $\epsilon\to0$, we have $\alpha\to(\log e)/I$, and
\begin{align*}
 & \sup_{P_{X,Y}:\,I(X;Y)=I}H^{*}(X\to Y)\\
 & >\left(\frac{I}{\log e}+\ln\frac{I}{\log e}+1-\gamma-12\frac{\log e}{I}\Big(\ln\frac{I}{\log e}\Big)^{2}\right)\log e\\
 & =I+\log I+(1-\gamma)\log e-\log\log e-12\frac{(\log e)^{2}}{I}\Big(\ln\frac{I}{\log e}\Big)^{2}.
\end{align*}

\end{IEEEproof}

\section{Acknowledgement}

This work was partially supported by a grant from the Research Grants Council of the Hong Kong Special Administrative Region, China {[}Project No: CUHK 14209823 (GRF){]}.

\section{AI Use Statement}

AI tools (GPT-6 Astra) were used to suggest improvements upon the previous result \cite{li2026discrete}, find the form of the bound in Theorem \ref{thm:PFR}, and find the key arguments used in the proofs of Proposition \ref{prop:properties}.\ref{enu:properties_bd}, Theorem \ref{thm:I_Jf}, Corollary \ref{cor:Jf_I}, Theorem \ref{thm:PFR} and Theorem \ref{thm:example}. All text in this paper are written by the author. The author verified all claims in this paper, and takes full responsibility for the originality and validity of the contents in this papers.

\bibliographystyle{IEEEtran}
\bibliography{ref}

\end{document}